\documentclass[fleqn,10pt,twocolumn]{wlscirep}
\usepackage[utf8]{inputenc}
\usepackage[T1]{fontenc}
\title{Temperature induced shifts of Yu-Shiba-Rusinov resonances in nanowire-based hybrid quantum dots}

\author[1]{Juan Carlos Estrada Salda\~{n}a}
\author[1,2]{Alexandros Vekris}
\author[1]{Victoria Sosnovtseva}
\author[1]{Thomas Kanne}
\author[1,3]{Peter Krogstrup}
\author[1]{Kasper Grove-Rasmussen}
\author[1,*]{Jesper Nyg{\aa}rd}
\affil[1]{Center for Quantum Devices, Niels Bohr Institute, University of Copenhagen, 2100 Copenhagen, Denmark}
\affil[2]{Sino-Danish Center for Education and Research (SDC)
SDC Building, Yanqihu Campus, University of Chinese Academy of Sciences,
380 Huaibeizhuang, Huairou District, 101408 Beijing, China}
\affil[3]{Microsoft Quantum Materials Lab Copenhagen, Niels Bohr Institute, University of Copenhagen, 2100 Copenhagen, Denmark}

\affil[*]{nygard@nbi.ku.dk}

\begin{abstract}
The strong coupling of a superconductor to a spinful quantum dot results in Yu-Shiba-Rusinov (YSR) discrete subgap excitations. In isolation and at zero temperature, the excitations are $\delta$ resonances. In transport experiments, however, they show as broad differential conductance peaks. We obtain the lineshape of the peaks and their temperature dependence in superconductor-quantum-dot-metal (S-QD-N) nanowire-based devices. Unexpectedly, we find that the peaks shift in energy with temperature, with the shift magnitude and sign depending on ground state parity and bias voltage. Additionally, we empirically find a power-law scaling of the peak area versus temperature. These observations are not explained by current models.

\end{abstract}
\begin{document}

\flushbottom
\maketitle

\thispagestyle{empty}


In a quantum dot-superconductor system, the exchange interaction of an unpaired, Coulomb-blockaded electron in the quantum dot with quasiparticles in the superconductor detaches discrete excitations from the edge of the superconducting gap~\cite{kirvsanskas2015yu}, as first explained by Yu, Shiba and Rusinov (YSR) for classical spins~\cite{yu1965,shiba1968classical,rusinov1969theory}.
When the coupling of the quantum dot to the superconductor is increased, the Kondo temperature, $T_K$, rises above the superconducting gap, $\Delta$, prompting a doublet$\to$singlet ground state transition marked by zero excitation energy~\cite{Satori1992Sep,deacon2010tunneling,lee2017scaling}. While the lineshape and temperature dependence of the normal-state spin-1/2 Kondo effect have been thoroughly characterized~\cite{vanderWiel2000Sep,Cronenwett1998Jul,Goldhaber-Gordon1998Dec,Nygard2000Nov}, its YSR superconducting analog is yet to be subjected to the same degree of scrutiny.

At finite temperature, the spectral weight of YSR excitations is characterized by an approximately Gaussian lineshape~\cite{Zitko2016May,Liu2019May}. In a realistic setup, in which the intrinsic superconductor-impurity system is probed by a scanning tunnelling tip~\cite{Yazdani1997Mar,Franke2011May,Kezilebieke2018Apr,Malavolti2018Dec,Ruby2018Apr,Cornils2017Nov} or a metallic contact~\cite{lee2014spin,jellinggaard2016tuning,deacon2010tunneling,grove2017yu,saldana2018supercurrent}, the excitations are measured as peaks in the differential conductance~\cite{Koerting2010Dec}, and various mechanisms may obscure their intrinsic lineshape. On one hand, the peaks can be dressed with a Lorentzian form in the presence of a relaxation channel for quasiparticles, which can be provided, for example, by a soft superconducting gap; i.e., a pseudogap populated by quasiparticle density of states up to the Fermi level~\cite{Martin2014Sep}. On the other hand, as by-product of the metallic lead, the normal-state spin-1/2 Kondo effect can emerge and distort the peak lineshape when $T<T^N_K$, where the superscript $N$ here is used to distinguish the Kondo temperature of the normal lead from $T_K$, the one of the superconducting lead, and $T$ is the temperature~\cite{Zitko2015Jan}. In addition, photon assisted tunnelling can broaden the superconducting density of states~\cite{Ast2016Oct,Pekola2010Jul}, though this issue may be solved by increasing the capacitance of the junction~\cite{Pekola2010Jul}. 

In planar semiconductor/superconductor devices, in which the gate tunability of the semiconductor is employed to define a quantum dot in close proximity to the superconductor~\cite{de2010hybrid}, a deteriorated interface between the superconductor and the semiconductor has been related to a soft superconducting gap~\cite{zhang2016ballistic,krogstrup2015epitaxy,Gazibegovic2017Aug,lee2012zero,Takei2013Apr}.
Earlier measurements of the temperature dependence of YSR excitations on soft-gapped devices reported no significant effects at $k_BT \ll \Delta$~\cite{Kumar2014Feb,Li2017Jan,Island2017Mar}. However, the use of a superconducting lead in place of a normal one led to non-equilibrium features at high temperatures~\cite{Kumar2014Feb,Li2017Jan,Island2017Mar}.

The interface improvement gained by the in-situ deposition of Al on InAs nanowires yields a hard gap -i.e., a gap devoid of quasiparticle density of states- in tunnel spectroscopy~\cite{krogstrup2015epitaxy,Chang2015Jan}. Using these nanowires, we define S-QD-N devices by either 1) etching Al~\cite{Chang2015Jan} or 2) shadowing in-situ~\cite{Gazibegovic2017Aug,Marnauza2020} to obtain a bare semiconductor channel. 
The devices are shown to have a hard gap, with $\Delta$ nearly temperature-independent in the temperature range explored.
At temperatures significantly smaller than $\Delta$, we observe a ground-state and bias-voltage dependent shift of the YSR subgap excitations, in apparent contradiction to recent calculations developed for the simpler S-QD system~\cite{Zitko2016May,Liu2019May}.  
The shift occurs irrespective of the conductance of the YSR peaks, implying a negligible role of the normal lead in this effect, and excluding a possibly lurking Kondo effect. 

\section*{Results}

\begin{figure*} [t!]
\centering
\includegraphics[width=0.85\linewidth]{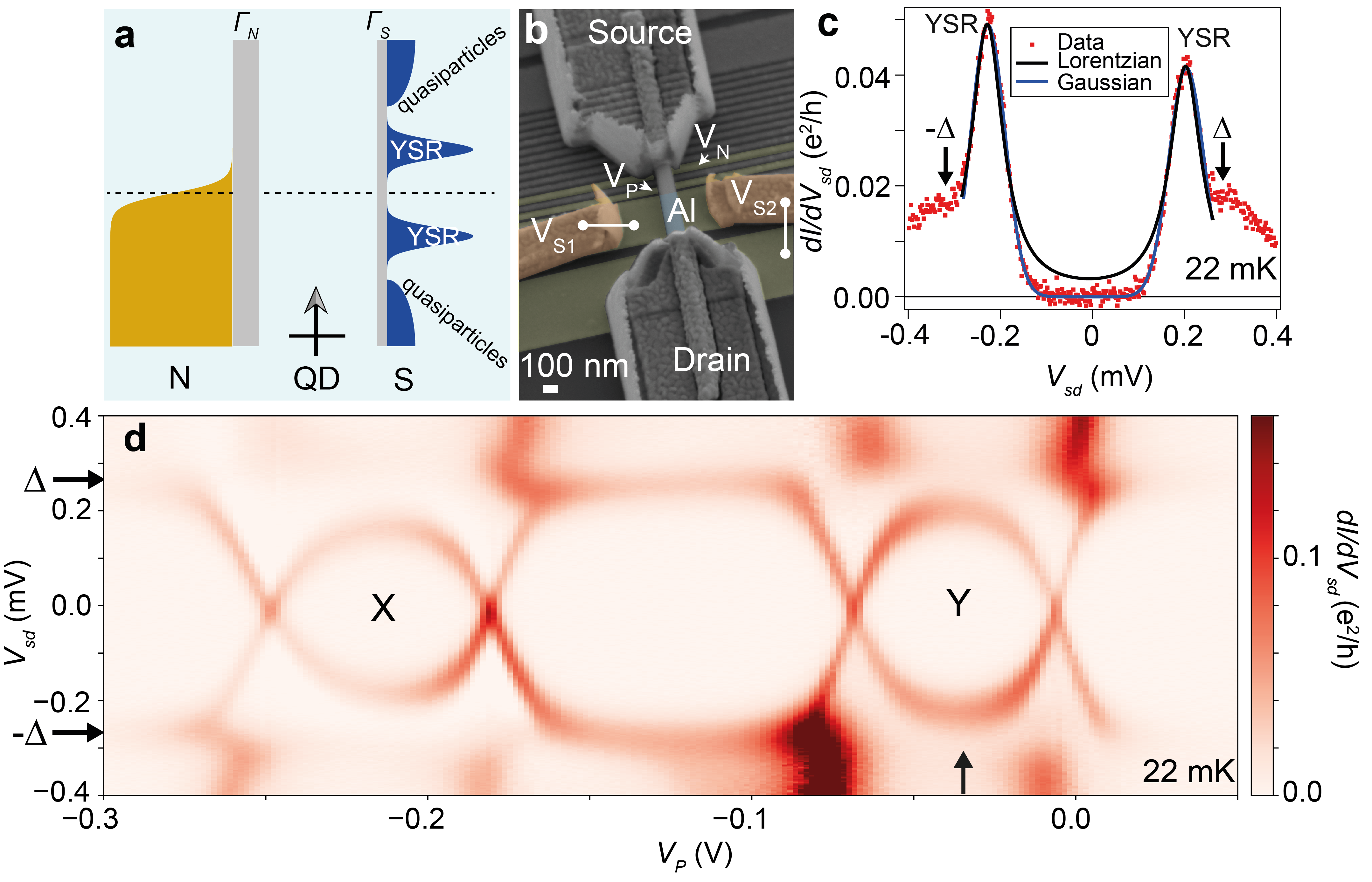}
\caption{\textbf{Gaussian YSR peaks.} (a) Sketch of the normal-quantum dot-superconductor (N-QD-S) system. (b) Scanning electron micrograph of the device. Bottom and side gates are false-colored in yellow and orange, respectively. Al appears in blue. (c) Fit of YSR peaks from differential conductance data to the sum of two Gaussian (Lorentzian) curves, shown in blue (black). $\Delta$ corresponds to the edge of the superconducting gap~\cite{Koerting2010Dec}, and agrees with the measured gap singularities (c.f.~Fig.~\ref{Fig9}a). (d) Colormap of YSR peaks vs.~plunger gate voltage, measured at $V_N=-6.82$ V, $V_{S1}=-6.56$ V, $V_{bg}=-20$ V. The color scale has been saturated to highlight the subgap features.}
\label{Fig1}
\end{figure*}

A sketch of the system in consideration is shown in Fig.~\ref{Fig1}a. From left to right, we show a normal metal lead with a Fermi-Dirac distribution at the electron temperature $T_e$, separated from a spinful quantum dot level by a tunnel barrier of coupling $\Gamma_N$. The exchange interaction of the spin $1/2$ with virtually excited quasiparticles in the superconductor via the barrier of coupling $\Gamma_S$ produces YSR $\delta$ resonances inside the gap, while the finite but low temperature leads to additional interaction with a small population of thermally fluctuated quasiparticles and produces broadened peaks at the $\delta$-peak position~\cite{Zitko2016May}. 
In our setup, the dilution refrigerator temperature $T$ is lower than $T_e$ (at base, $T=20-30$ mK while $T_e \approx 80$ mK), as it is typically the case~\cite{Torresani2013Dec,Feshchenko2015Sep}. A doped-Si substrate backgate increases the lead capacitance ($C\approx10$ pF) of our devices, which has been shown to reduce environmentally-assisted tunnelling \cite{Pekola2010Jul}. The backgate, $V_{bg}$, is also used as an additional tuning knob of the quantum dots. Al is covering three facets of the nanowire, and Au is used as contact to the bare facets. 
In Methods, we show details of device fabrication and evidence of hard gap in our devices. The differential conductance, $dI/dV_{sd}$, of the devices is measured with a lock-in amplifier, where $V_{sd}$ is the source-drain bias voltage.   

We first focus on the device in which Al was etched, shown in Fig.~\ref{Fig1}b. Bottom gate $V_N$ controls the coupling of the QD to Au, while bottom and side gates $V_{S1}$ and $V_{S2}$ control its coupling to Al. $V_{S2}$ was kept at -5.5 V throughout the experiment. Bottom gate $V_P$ acts as QD plunger gate, controlling the charge occupation.

\begin{figure*} [t!]
\centering
\includegraphics[width=1\linewidth]{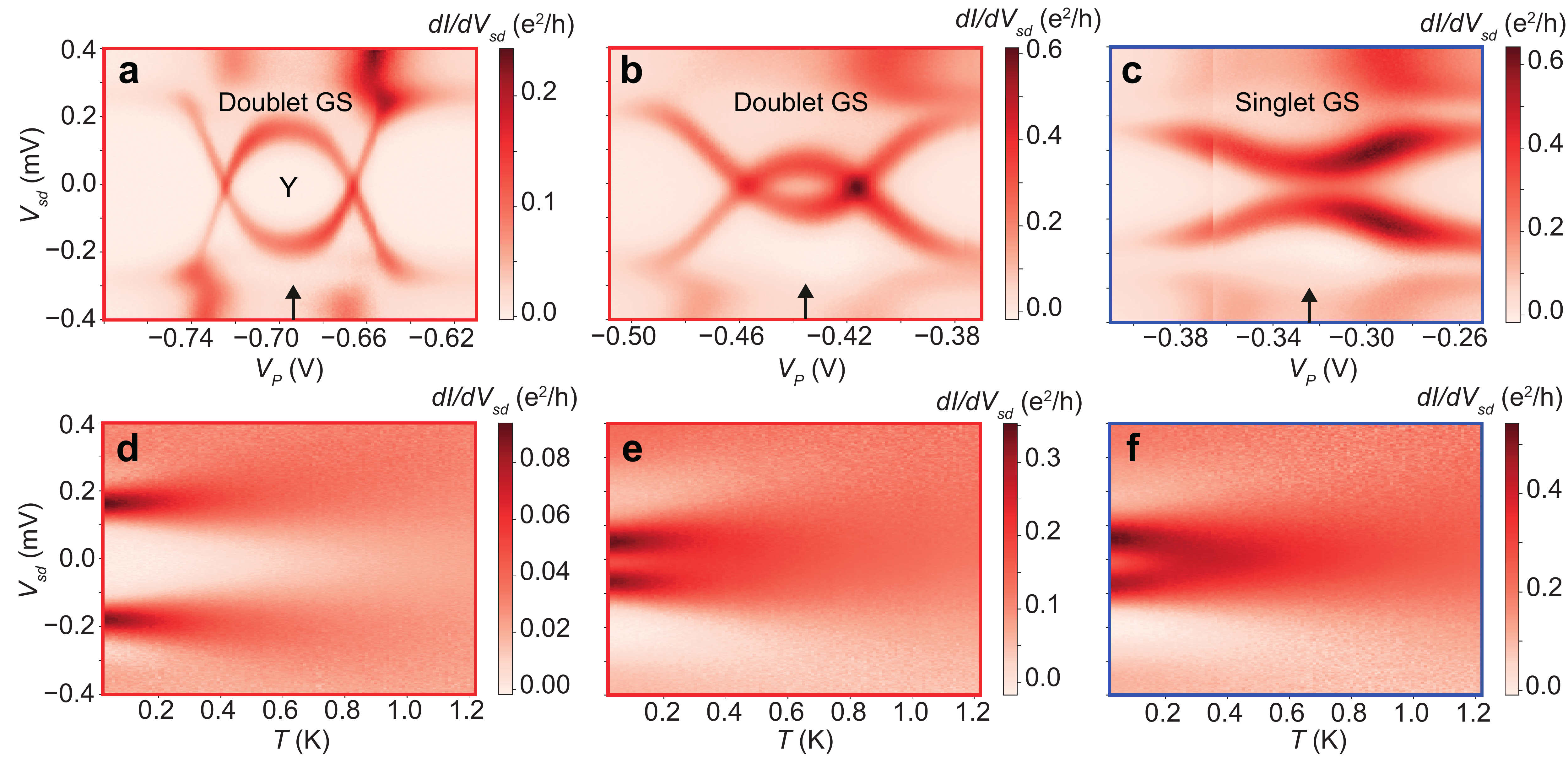}
\caption{\textbf{Tuning YSR states across a doublet-singlet ground state transition.} (a-c) Colormaps of YSR peaks in charge state Y at increasing coupling to the superconducting lead. (d-f) Temperature dependence of YSR peaks at the e-h symmetry point of maps (a-c), indicated by an arrow. (a) $V_{S1}=-6.42$ V, (b) $V_{S1}=-6.76$ V, (c) $V_{S1}=-6.94$ V. $V_P$ is compensated for the change in $V_{S1}$. $V_N$ and $V_{bg}$ were kept at -6.82 V and -20 V, respectively.}
\label{Fig2}
\end{figure*}

\begin{figure*} [t!]
\centering
\includegraphics[width=1\linewidth]{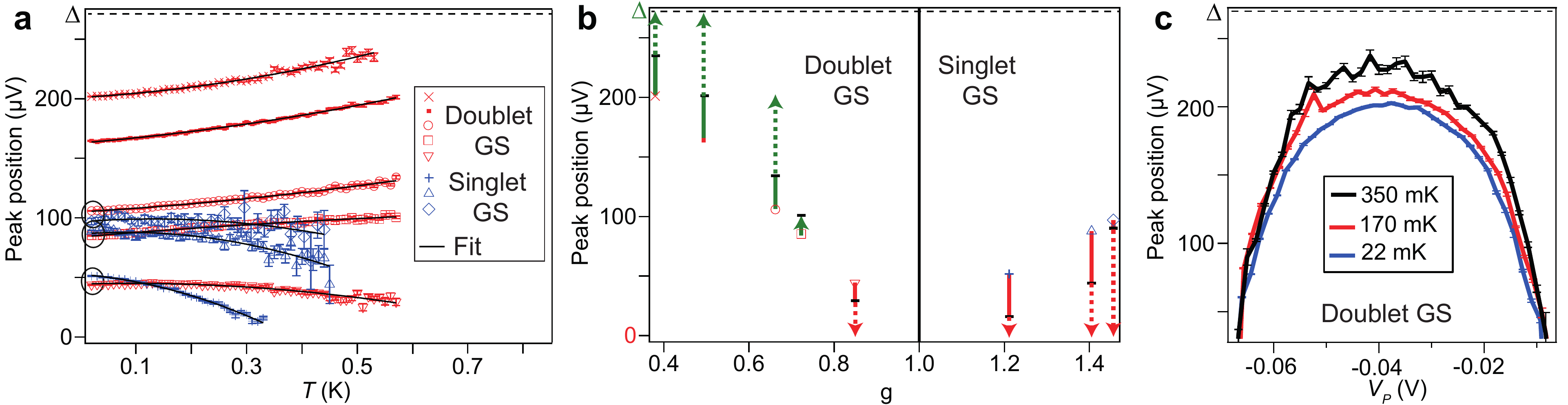}
\caption{\textbf{Temperature dependence of extracted bias position of YSR peaks in charge state Y.} (a) Temperature dependence of peak position at the e-h symmetry point across the doublet (red) $\to$ singlet (blue) ground state (GS) transition. (b) Peak position corresponding to the lowest and highest temperature data points in each of the datasets from (a) vs.~ $g \sim J$, the exchange coupling. Shifts towards the points $V_{sd}=\Delta$ ($V_{sd}=0$) are indicated by green (red) arrows. Solid lines end at the temperature above which the Gaussian fit cannot be reliably carried out, corresponding to the highest temperature shown in (a) for each dataset. The arrows are extended beyond that as dotted lines up to the final position of the YSR peaks at $1.2$ K. (c) Peak position vs. plunger gate voltage at different temperatures.}
\label{Fig3}
\end{figure*}

In our setup, YSR $dI/dV_{sd}$ peaks present in $V_{sd}-dI/dV_{sd}$ traces can be fitted to the sum of two Gaussian lineshapes over a range of gate voltages and temperatures. This allows us to extract values for the position, height, and full-width at half-maximum (FWHM) of the peaks against these variables. 
Figure \ref{Fig1}c shows a trace exemplifying the fits in comparison to the sum of two Lorentzians. The tails of the Lorentzians drop too slowly to account for the data around zero bias, an effect observed for all the fitted data.

Figure \ref{Fig1}d shows a map of subgap $dI/dV_{sd}$ as function of $V_{sd}$ and plunger gate voltage, $V_P$. The two small loops identified by X and Y correspond to YSR doublet$\to$singlet excitations. We independently corroborate their doublet ground state nature through their evolution in an external magnetic field~\cite{lee2014spin,jellinggaard2016tuning} (see Methods). The charging energies corresponding to these spinful charge states are $U=3.1$ meV and $U=2.7$ meV, respectively, obtained from Coulomb-diamond spectroscopy, whereas the gap singularities appear at $|\Delta|=0.27$ meV. The condition $U \gg \Delta$ places the system within the YSR regime~\cite{kirvsanskas2015yu}. The trace in Fig.~\ref{Fig1}c was acquired at the electron-hole (e-h) symmetry point of charge state Y, indicated by a vertical arrow in Fig.~\ref{Fig1}d.

Figures \ref{Fig2}a-c show three examples of the effect of changing $V_{S1}$ for charge state Y. From plots \ref{Fig2}a to \ref{Fig2}c, the YSR loop shrinks and opens again as the same ground state changes from doublet to singlet via gate control of $\Gamma_S$~\cite{lee2014spin,jellinggaard2016tuning}. Figures \ref{Fig2}d-f show in turn the temperature dependence of the YSR peaks at the e-h symmetry point of the respective colormaps in Figs.~\ref{Fig2}a-c. As the temperature increases, the pair of peaks which corresponds to doublet ground state and is closer to $\Delta$ in low-temperature bias position splits apart (Fig.~\ref{Fig2}d).
Strikingly, when the initial bias position of the peaks is roughly the same, as in Figs.~\ref{Fig2}e and \ref{Fig2}f, the pair of peaks which corresponds to singlet ground state (Fig.~\ref{Fig2}f) goes faster towards zero-bias than the pair which corresponds to doublet ground state (Fig.~\ref{Fig2}e). In contrast, Numerical Renormalization Group (NRG) calculations of the spectral weight of YSR peaks in the single-impurity Anderson model with a conventional superconducting lead have predicted a temperature-independent peak position for a constant gap~\cite{Zitko2016May}, and the opposite behavior to our observations for a temperature-dependent $\Delta$~\cite{Zitko2016May,Liu2019May}. To the best of our knowledge, no current models account for this behavior.


To obtain a quantitative description of the variation of the peak position against temperature, we fitted to a Gaussian the YSR peak at positive bias from the three datasets in Figs.~\ref{Fig2}d-f and from five more datasets taken at intermediate peak positions; all of them at the e-h symmetry point of charge state Y. The temperature range covered by the fit (from 22 mK to $\approx 550$ mK, or from $0.01\Delta$ to $\approx 0.25\Delta$) corresponds to the low-temperature regime~\cite{Zitko2016May}, where $\Delta$ is constant ($T \ll T_c=2.2$ K) and significant quasiparticle thermal excitation is not expected to occur. At 22 mK (550 mK), the quasiparticle density in the Al lead is theoretically estimated as $D(E_F)\sqrt{2\pi \Delta k_BT}exp(-\Delta/k_BT)\sim 10^{-65}$ nm$^{-3}$ ($\sim 10^{-5}$ nm$^{-3}$), where $D(E_F)=23$ eV$^{-1}$nm$^{-3}$ is the density of states of Al at the Fermi energy~\cite{Higginbotham2015Sep,Saira2012Jan,Wilson2001Jul}. We observe, however, that the subgap conductance increases with temperature, indicating non-negligible quasiparticle thermal excitation (see Fig.~\ref{Fig10}f under Methods). 

Figure \ref{Fig3}a shows the extracted evolution in temperature of the position of the peaks, five of which correspond to doublet ground state (in red) and three to singlet ground state (in blue). As elsewhere in this work, error bars correspond to standard deviation. The datasets have been fitted to parabolas $y=a_0+a_1T+a_2T^2$ (solid lines) in order to indicate that they do not change faster than $T^2$. 
Black circles pair datasets of singlet and doublet ground states whose initial bias position roughly match. The qualitative picture extracted from Fig.~\ref{Fig2} is corroborated for such pairs; namely, when having approximately the same initial bias position, datasets of singlet ground state shift faster towards zero bias than datasets of doublet ground state. In addition, a new detail is worth mentioning. The curvature of the datasets of doublet ground state changes from positive to negative as $V_{sd} \to 0$; i.e., as the peaks are biased away from $\Delta$. In the case of the datasets of singlet ground state, the curvature becomes more negative as $V_{sd} \to 0$ for the data available. 

This ground-state and bias-position dependent behavior can be parametrized by the exchange coupling, $J$. In Fig.~\ref{Fig3}b, we plot the endpoints of each dataset in Fig.~\ref{Fig3}a as a function of $g=\pi JS D(E_F)$, where $S$ is the spin. We convert YSR peak position to $g$ using $E_{YSR} = \Delta \frac{1-g^2}{1+g^2}$, valid in the classical spin limit \cite{Zitko2015Jan,kirvsanskas2015yu}. YSR peaks of doublet ground state whose low-temperature bias position is closer to $\Delta$, corresponding to small $g$, shift towards $\Delta$ as the temperature is increased. The shift direction is reversed when $g$ is tuned towards the doublet-singlet ground state transition. The reversal occurs between $g=0.7-0.85$, when the YSR peak of doublet ground state of lowest bias position shifts towards zero bias. When $g=1.2-1.5$, after the transition occurs and the ground state is a singlet, the remaining datasets shift towards zero bias. 

In Fig.~\ref{Fig3}c, the YSR peak position across charge state Y is shown to shift towards larger bias with temperature, indicating that the observed temperature-shifting behavior is not exclusive of the e-h symmetry point. The dataset used to extract the plunger gate dependence of the peak position at various temperatures is shown under Methods in Fig.~\ref{Fig10}. 


\begin{figure*} [t!]
\centering
\includegraphics[width=1\linewidth]{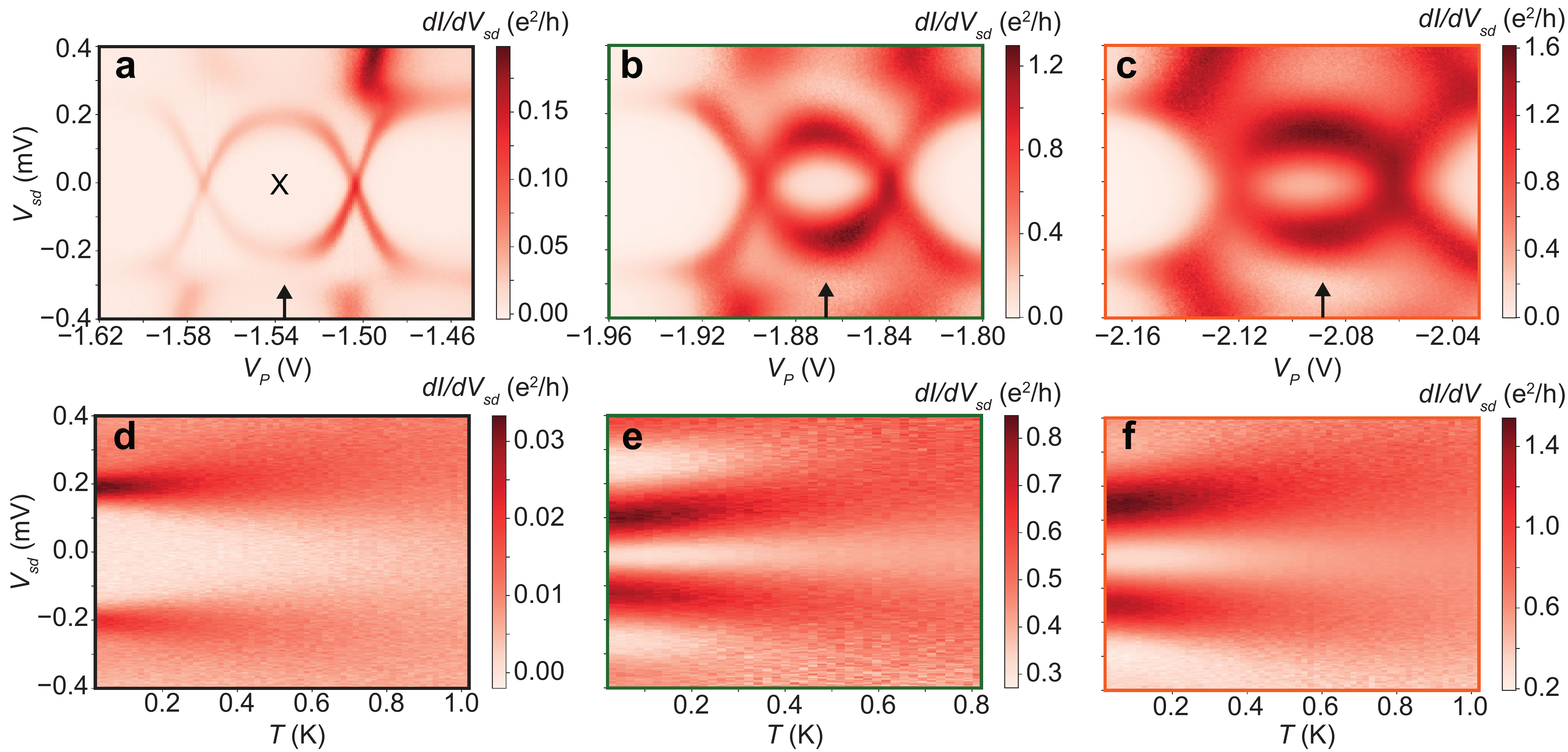}
\caption{\textbf{Tuning the magnitude of YSR peaks.} (a-c) Colormaps of YSR peaks in charge state X (of doublet ground state) at increasing peak conductance. (d-f) Temperature dependence of YSR peaks at the e-h symmetry point of maps (a-c), indicated by an arrow. (a) $V_N=-6.77$ V, (b) $V_N=-6.22$ V, (c) $V_N=-5.7$ V. $V_P$ is compensated for the change in $V_{N}$. $V_{S1}$ and $V_{bg}$ were kept at -6.4 V and -11.55 V, respectively.}
\label{Fig4}
\end{figure*}

\begin{figure*} [t!]
\centering
\includegraphics[width=0.9\linewidth]{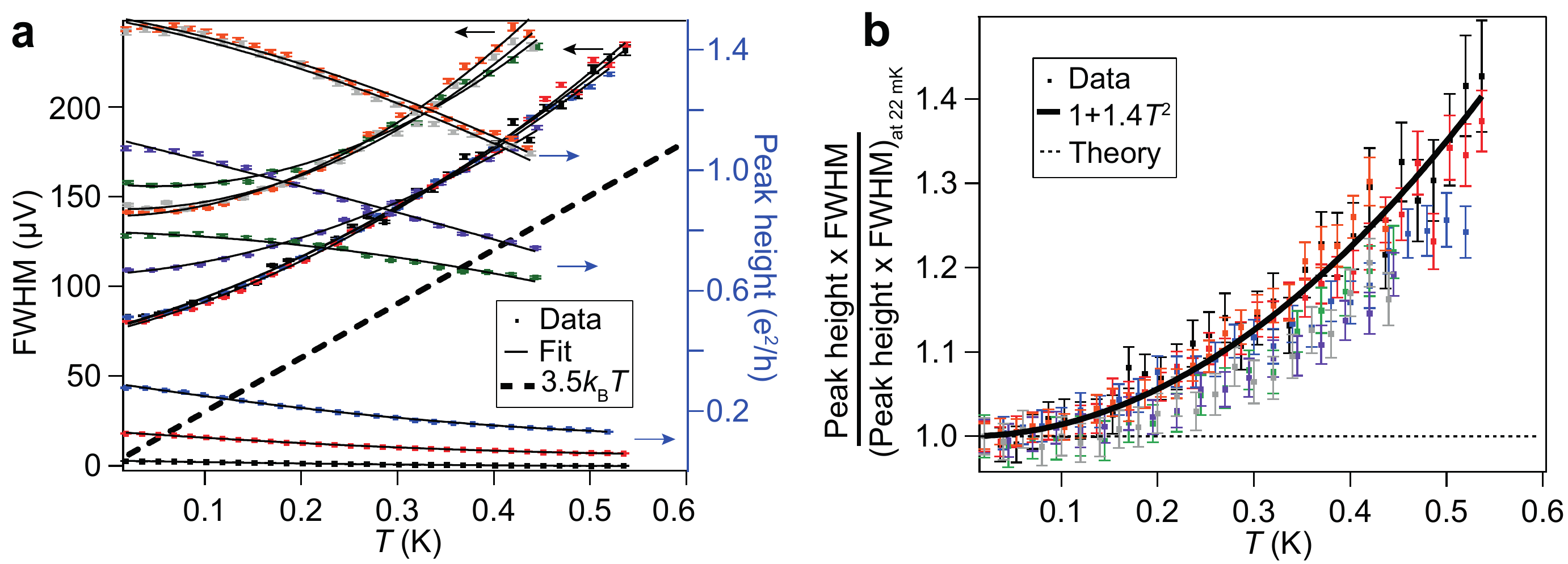}
\caption{\textbf{Temperature dependence of YSR peaks when tuned to different magnitude.} (a) Extracted peak height (right axis) and FWHM (left axis) vs.~temperature at the e-h symmetry point of charge state X of doublet ground state. (b) Temperature dependence of the product of peak height and FWHM obtained from (a) normalized by their values at 22 mK. Datasets from (a) and (b) are color-coded in the same way.}
\label{Fig5}
\end{figure*}

\begin{figure*} [t!]
\includegraphics[width=1\linewidth]{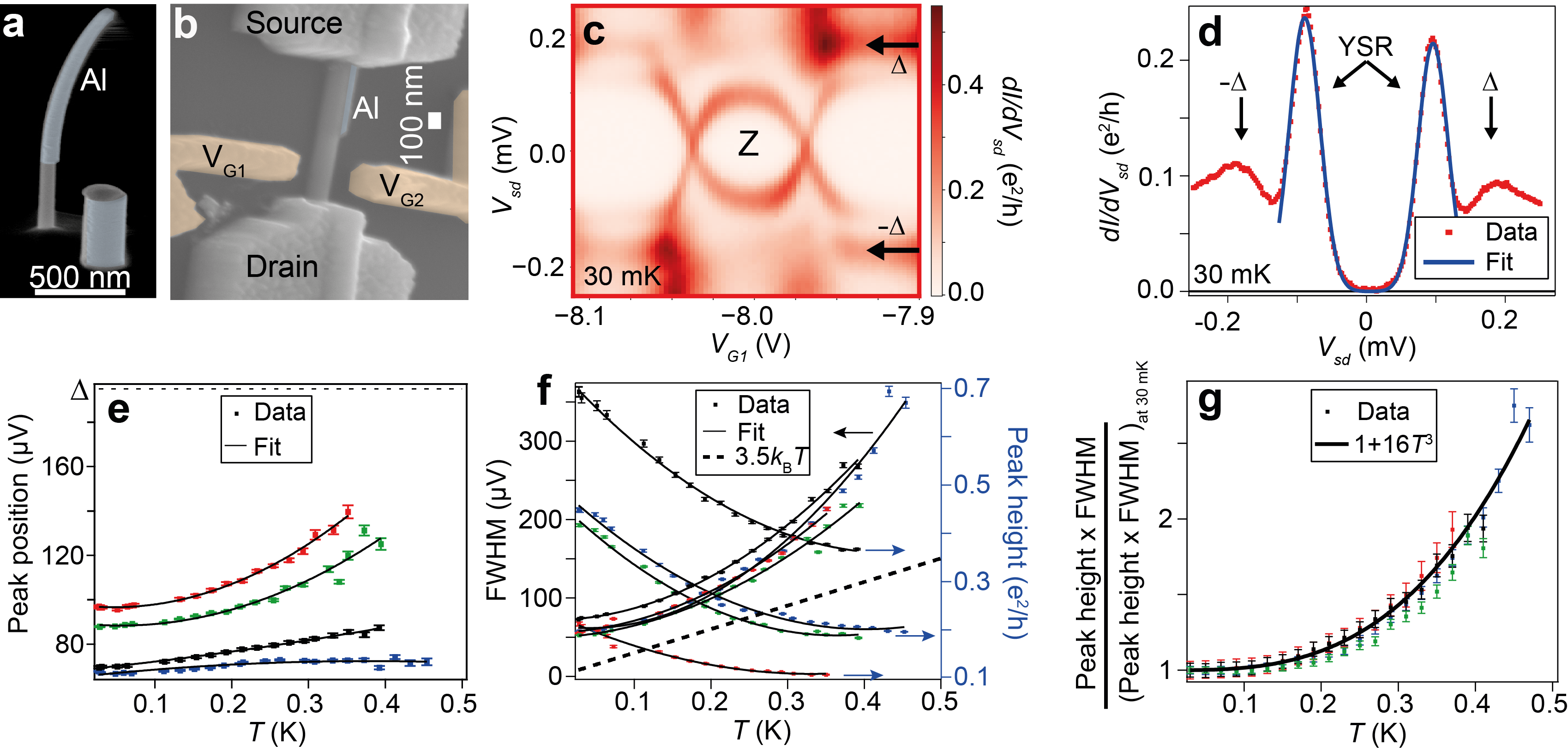}
\centering
\caption{\textbf{Data from additional device.} (a,b) Scanning electron micrograph of (a) a typical set of as-grown wires and (b) the device. (c) Colormap of YSR subgap peaks evolving in plunger gate voltage across spinful charge state Z. Arrows indicate the position of the gap singularities. (d) Fit to Gaussians of YSR peaks at the e-h symmetry point of Z. (e,f) Temperature dependence of (e) peak position and (f) peak height, FWHM extracted from fitting YSR peaks at the e-h symmetry point of four different charge states of doublet ground state, including Z. All datasets have been fitted to parabolas of the form $y=a_0+a_1T+a_2T^2$, to indicate that they do not change faster than $T^2$. (g) Temperature dependence of the product of peak height and FWHM scaled by their values at 30 mK. The four datasets collapse in a single cubic curve.}
\label{Fig6}
\end{figure*}

We now turn our attention to the peak height and width dependence in temperature against an increase of the coupling of the quantum dot to the normal lead. We plot in Figs.~\ref{Fig4}a-c three examples of the effect of changing $V_N$ within charge state X in the doublet ground state. From left to right plots, the conductance of the subgap states is significantly enhanced. 
We interpret the enhancement of the conductance as stemming from a decrease in the $\Gamma_N$, $\Gamma_S$ asymmetry $r$ from $\Gamma_N \ll \Gamma_S$ to $\Gamma_N \sim \Gamma_S$ due to an increase of $\Gamma_N$. We can offer an order of magnitude of this asymmetry from the relation peak height$=2e^2/h \times 4r/(1+r)^2$, where $r=\Gamma_N/ \Gamma_S$~\cite{Meir1992Apr,lee2014spin}. By fitting YSR peaks with Gaussians at the e-h symmetry point of charge states X and Y across the entire gate space ($V_N$, $V_{S1}$) explored, we obtain a peak height range of $0.003e^2/h \to 1.5e^2/h$, or $\Gamma_N / \Gamma_S \sim 8\times 10^{-4}\Gamma_S \to 0.3$. 

Figures \ref{Fig4}d-f show colormaps of the temperature dependence of YSR peaks at the e-h symmetry point of each of the examples from Figs.~\ref{Fig4}a-c. In confirmation of our previous observation, the YSR peaks, which are of doublet ground state and away from zero bias, split apart as the temperature is increased.
We highlight another observation; as the temperature rises the peak height is decreased and the peak width is increased.

In Fig.~\ref{Fig5} we summarize quantitative data on this effect. Figure \ref{Fig5}a shows a plot of decreasing peak height curves and increasing FWHM curves as temperature is increased. These were acquired from a fit of the positive-bias YSR peak in the three datasets of Figs.~\ref{Fig4}d-f plus four additional datasets, all taken at the e-h symmetry point of charge state X of doublet ground state. All curves were fitted to parabolas $y=a_0+a_1T+a_2T^2$, to indicate that they do not change faster than $T^2$. For comparison, we plot the linear broadening $3.5 k_BT$ due to the Fermi-Dirac distribution of the normal lead. The slope of the FWHM data is smaller than $3.5 k_BT$ below 0.2 K, and larger than $3.5k_BT$ above 0.4 K, while its magnitude is larger than $3.5k_BT$, indicating that thermal broadening by the metallic lead is not the only broadening mechanism. Note that additional thermal broadening is expected in the S-QD side even in the absence of a metallic lead~\cite{Zitko2016May}, whereas the tunnel coupling to the normal lead at zero temperature can broaden the peaks even further~\cite{kirvsanskas2015yu,Koerting2010Dec}. 

While the upper bound of the theoretical conductance of YSR peaks is $2e^2/h$~\cite{Martin2014Sep}, in the presence of a finite quasiparticle relaxation tunnelling rate to a continuum of states, an increase of relaxation rate or temperature leads to a decrease in peak conductance as $\sim 1/(\Gamma+T)$, where the lifetime $\Gamma$ includes the relaxation rate~\cite{grove2017yu,Martin2014Sep}. In the same formalism, the FWHM scales as $\sim (\Gamma+T)$. Therefore, the product of FWHM and peak height, which provides the area of the peak, is a constant independent of temperature. The constant is equal to 1 if the product is normalized by its value at $T=0$. Surprisingly, the products of the peak height and FWHM of the seven datasets in Fig~\ref{Fig5}a scaled by their values at 22 mK bunch into a single quadratic curve, as shown in Fig.~\ref{Fig5}b. This occurs despite of widespread change in peak height (of about 2 orders of magnitude), FWHM (from $0.3\Delta$ to $0.5\Delta$) and peak position (from $0.3\Delta$ to $0.7\Delta$). For comparison, a constant dashed line equal to 1, predicted by the relaxation formalism, is also shown.

Finally, we report results from a second device fabricated from a nanowire shadowed during in-situ Al deposition by a thick and shorter nanowire~\cite{Marnauza2020}. The resulting Al/nanowire heterostructure, shown in \ref{Fig6}a, eliminates the need to etch Al to form the junction. Figure \ref{Fig6}b shows a scanning electron micrograph of the device. Side gate $V_{G1}$ was used as plunger gate, while side gate $V_{G2}$ and a substrate backgate $V_{bg}$ were used to bring the wire close to charge depletion. YSR excitations of doublet ground state form loops identified by their smaller size than their adjacent counterparts, as exemplified in the $dI/dV_{sd} (V_{sd},V_{G1})$ map of Figure \ref{Fig6}c. From Coulomb-diamond spectroscopy, we determined the charging energy of the associated spinful charge state indicated by Z as $U=1.1$ meV, and $\Delta = 0.195$ meV. As in the previous device, the sum of two Gaussians fits YSR peaks in $dI/dV_{sd} (V_{sd})$ traces up to the gap edge. An instance of this is shown in Fig.~\ref{Fig6}d.

In Figs.~\ref{Fig6}e,f, we show the peak position, height and FWHM extracted from fitting the temperature dependence of four charge states of doublet ground state at their e-h symmetry points, including state Z. The qualitative similarities of the data in both devices is noticeable. As before, the peak-position datasets in Fig.~\ref{Fig6}e exhibit a bias-dependent change of curvature with temperature. Similarly, in Fig.~\ref{Fig6}f the FWHM and peak height vs.~temperature datasets obey opposite trends, while the FWHM shows a curvature increase with respect to $3.5 k_BT$. 

Nonetheless, there is a quantitative difference. In Fig.~\ref{Fig6}g, we plot the product of peak height and FWHM normalized by their values at 30 mK, the lowest temperature at which data was recorded for this device. The four datasets collapse into the same curve, resembling the result from the previous device. However, the curve into which they collapse grows with $T^3$, whereas that of the previous device grew with $T^2$.

\section*{Discussion}

As commented above, a Kondo singlet with Kondo temperature $T^N_K$ can form with the normal lead in the QD-N part of the S-QD-N system~\cite{Zitko2015Jan}.
To address this possibility, we estimate the temperature of the Kondo resonance in the isolated N-QD system. $T^N_K$ then depends on gate voltage through the level position, $\epsilon_0$, of the QD as $T^N_K=\frac{\sqrt{\Gamma U}}{2}e^{\pi \epsilon_0 (\epsilon_0 +U)/\Gamma U}$, where $\Gamma$ is the linewidth of the level~\cite{Goldhaber-Gordon1998Dec}. Due to the sensitivity of the exponent to changes in $\Gamma=\Gamma_N$, a small variation in $\Gamma_N$ at the e-h symmetry point results in a large change in $T^N_K$. At the singlet-doublet transition point~\cite{Satori1992Sep}, $k_BT_K=0.3\Delta=81$ $\mu$eV for $U=3.1$ meV, which we can use to estimate $\Gamma_S=1$ meV and an upper bound of $T^N_K \approx 1.7$ mK for the YSR peaks of doublet ground state of largest conductance ($r=0.3$). Such a small $T^N_K$ indicates that the Kondo effect of the normal lead is not playing an important role at the e-h symmetry point.

In view of this, the role of the N-QD part of the system is reduced to a non-perturbing tunnel probe, and an explanation for the ground-state and peak-position dependent YSR peak behavior against temperature is to be found in the remaining part of the system, QD-S. A trivial reduction of $\Gamma_S$ with temperature, which would increase monotonically the energy of the singlet state, is ruled out based on the non-monotonic behavior of the curvature of the peak-position datasets depending on initial $V_{sd}$ position. In turn, the $f(T)=T^\alpha$ dependence extracted from the normalization of the peak area has a less evident origin. Phonon-mediated quasiparticle relaxation could in principle provide temperature-dependent broadening in our N-QD-S setup~\cite{Kozorezov2008Nov,Hijmering2009Jun}, but has so far only been used phenomenologically in analyzing the outcome of the S-YSR setup, in which an additional superconductor probes YSR excitations, leading to the need of deconvolving intrinsic YSR effects from those of the superconducting probe~\cite{Ruby2015Aug}.

In the presence of a significant relaxation tunnelling rate from YSR subgap resonances to a continuum of states, an asymmetry of the height of the peaks in positive-negative bias voltage is expected~\cite{Martin2014Sep}. However, this additional tunnelling rate theoretically results in a Lorentzian YSR peak lineshape~\cite{Martin2014Sep}, while we observe Gaussian lineshapes. In addition, the temperature dependence of the YSR peak area remarkably deviates from the expected dependence given by the relaxation model presented in Refs.~\cite{Martin2014Sep,grove2017yu}. Despite these inconsistencies, the height of the YSR conductance peaks is asymmetric in bias voltage even at the e-h symmetry point, as it is readily seen from Figs.~\ref{Fig1} and \ref{Fig6}, in apparent agreement with an important relaxation tunnelling rate~\cite{Martin2014Sep}. It is unclear which of these observations are determinant arguments in favor or against the existence of a finite relaxation tunnelling rate in our devices.

\begin{figure*} [t!]
\centering
\includegraphics[width=1\linewidth]{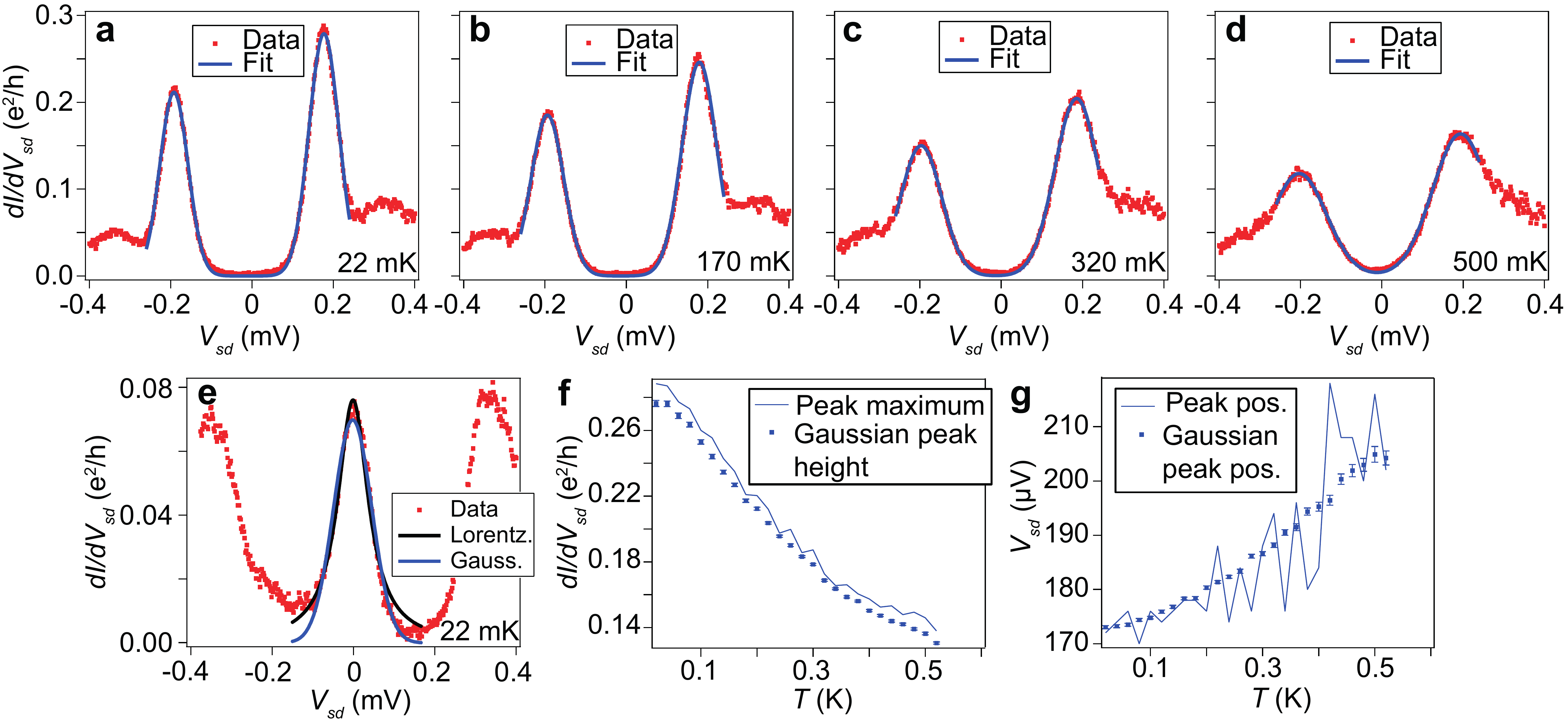}
\caption{\textbf{Details of the fit of YSR peaks to Gaussian curves.} (a-d) Robustness of the fit against temperature. (e) Zero-bias differential conductance peak obtained from the crossing of two YSR peaks at the right doublet-singlet crossing of state Y in Fig.~\ref{Fig1}d (at $V_P=-0.006$ V) fitted to a single Lorentzian (black) and Gaussian (blue) curves. While the Gaussian curve captures better the tail at positive bias of the crossed YSR peaks, it fails to do so at negative bias due to the presence of the superconducting gap edge. (f) Peak maximum and fitted peak height against temperature. (g) Position of peak maximum and fitted peak position against temperature.}
\label{Fig7}
\end{figure*}

Majorana zero-modes, which can arise in Rashba semiconductor nanowires coupled to superconductors under a properly oriented magnetic field, can give rise to zero-bias differential conductance peaks in tunnel spectroscopy~\cite{deng2016majorana,Zhang2018Mar}. Nevertheless, peaks from YSR states bear distinctive features from those from Majorana modes. While the lineshape of Majorana peaks is a Lorentzian~\cite{Zhang2018Mar,Zazunov2016Jul}, our data indicates that well-separated YSR peaks have a Gaussian lineshape. Note, however, that when two YSR peaks collapse into a single zero-bias peak (e.g., at a singlet-doublet transition), the Lorentzian lineshape might be harder to rule out (c.f.~Fig.~\ref{Fig7}e). Additional differences can be found in the change of their FWHM with temperature in comparison to $3.5 k_BT$~\cite{Zhang2018Mar}. 

To summarize, we have simplified the complex S-QD-N system by employing a hard superconducting gap and non-invasive N probes, while exploiting the gate tunability of YSR excitation and ground state energies. We have extensively characterized the temperature dependence of YSR resonances in two devices, establishing a basis for further experimental and theoretical work. In particular, the origin of shifts in bias voltage of the resonances against temperature is not explained by current models~\cite{Zitko2016May,Liu2019May}.

\section*{Methods}

\begin{figure*} [t!]
\centering
\includegraphics[width=0.7\linewidth]{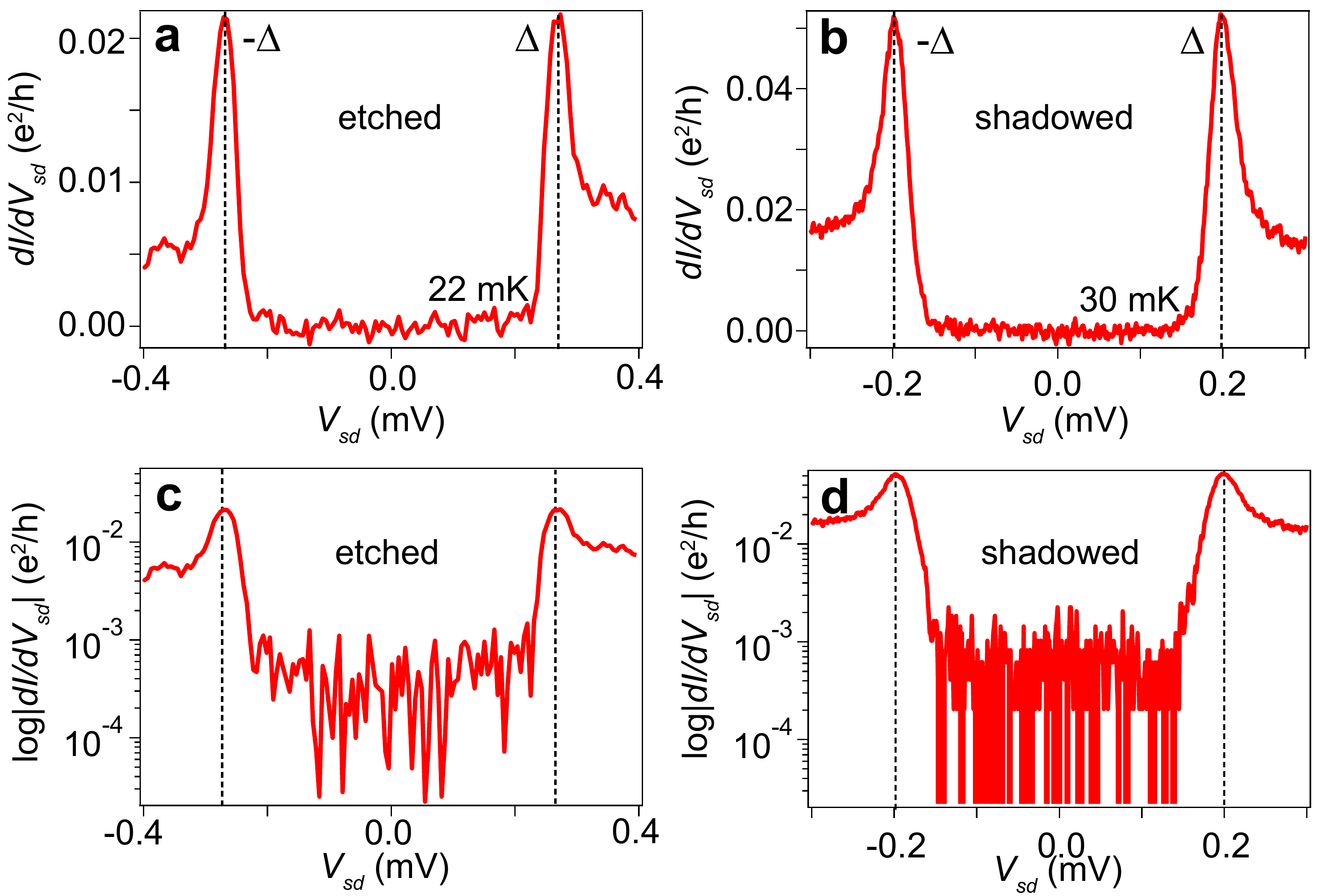}
\caption{\textbf{Superconducting gap.} $dI/dV_{sd}(V_{sd})$ plots in (a,b) linear and (c,d) logarithmic scale of the superconducting gap in the (a,c) etched and (b,d) shadowed devices. The gap singularities are identified by dashed lines. }
\label{Fig9}
\end{figure*}

Below we provide additional details of the fit and of device fabrication, as well as an independent corroboration of the ground state of the S-QD-N system.

\subsection*{Fabrication of the devices}

To fabricate the first device, a 110-nm wide InAs epitaxially half-shell Al-covered nanowire was deterministically deposited on a bed of local bottom gates and additional side-gates were defined during contact deposition. By etching-off the 7-nm-thick Al from the top half of the nanowire and contacting the resulting bare wire with Ti/Au, a N-QD-S junction was defined with a 250 nm channel of bare wire. The Au contact on the Al-lead side was 400 nm away from the channel. The leads ended in large-area bonding pads with a capacitance of $\approx10$ pF estimated by a simple parallel-plate capacitor model to the Si backgate through 200 nm of Si oxide.

To fabricate the second device, a shadowed 90-nm wide InAs wire was deterministically deposited on a Si substrate of similar characteristics as the previous device, leading to similar leads capacitance. Side-gates were defined during evaporation of the ohmic contacts to the bare wire. The bare-wire channel between the 20-nm thick Al film and the Ti/Au contact was 450 nm long. The Au contact on the Al-lead side was 500 nm away from the channel.     

\subsection*{Measurements}

The devices were voltage-biased. The DC current was acquired using a digital multimeter, while the $dI/dV_{sd}$ signal was recorded using standard lock-in amplifier techniques. To obtain $dI/dV_{sd}$ an excitation of 3 $\mu$V on top of $V_{sd}$ was applied at a frequency of 116.69 Hz for the device in which Al was etched, and of 132.7 Hz for the device in which Al was shadowed.

\subsection*{Details of the fit}

\begin{figure*} [t!]
\centering
\includegraphics[width=1\linewidth]{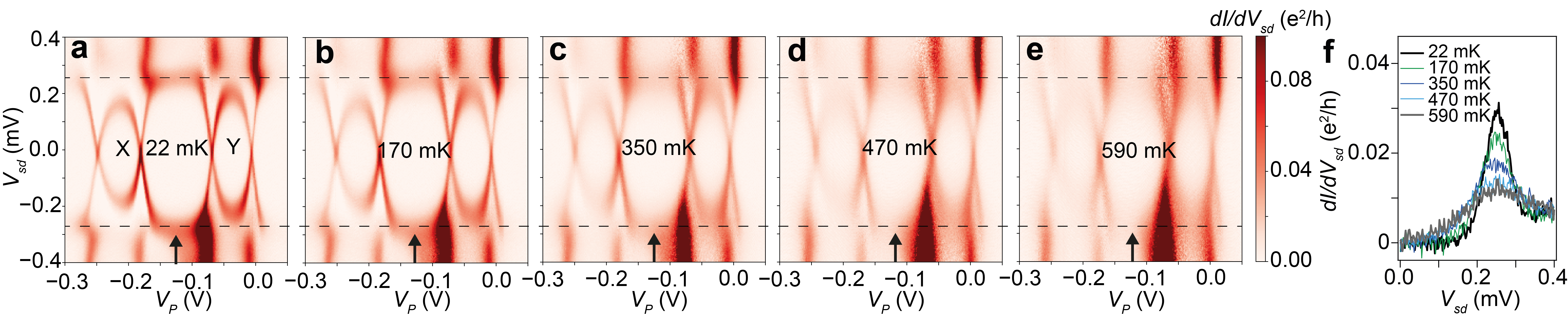}
\caption{\textbf{Variation of the gap against temperature.} (a-e) Colormaps of the evolution of Fig.~\ref{Fig1}d in temperature. The dashed lines indicate the position of the YSR peaks in the charge state of singlet ground state between charge states X and Y, which are related to the edges of the gap~\cite{Koerting2010Dec}. These do not move with temperature. (f) Positive-bias linecuts through the center of the even singlet sector, representing the temperature dependence of the gap.}
\label{Fig10}
\end{figure*}

\begin{figure*} [t!]
\centering
\includegraphics[width=1\linewidth]{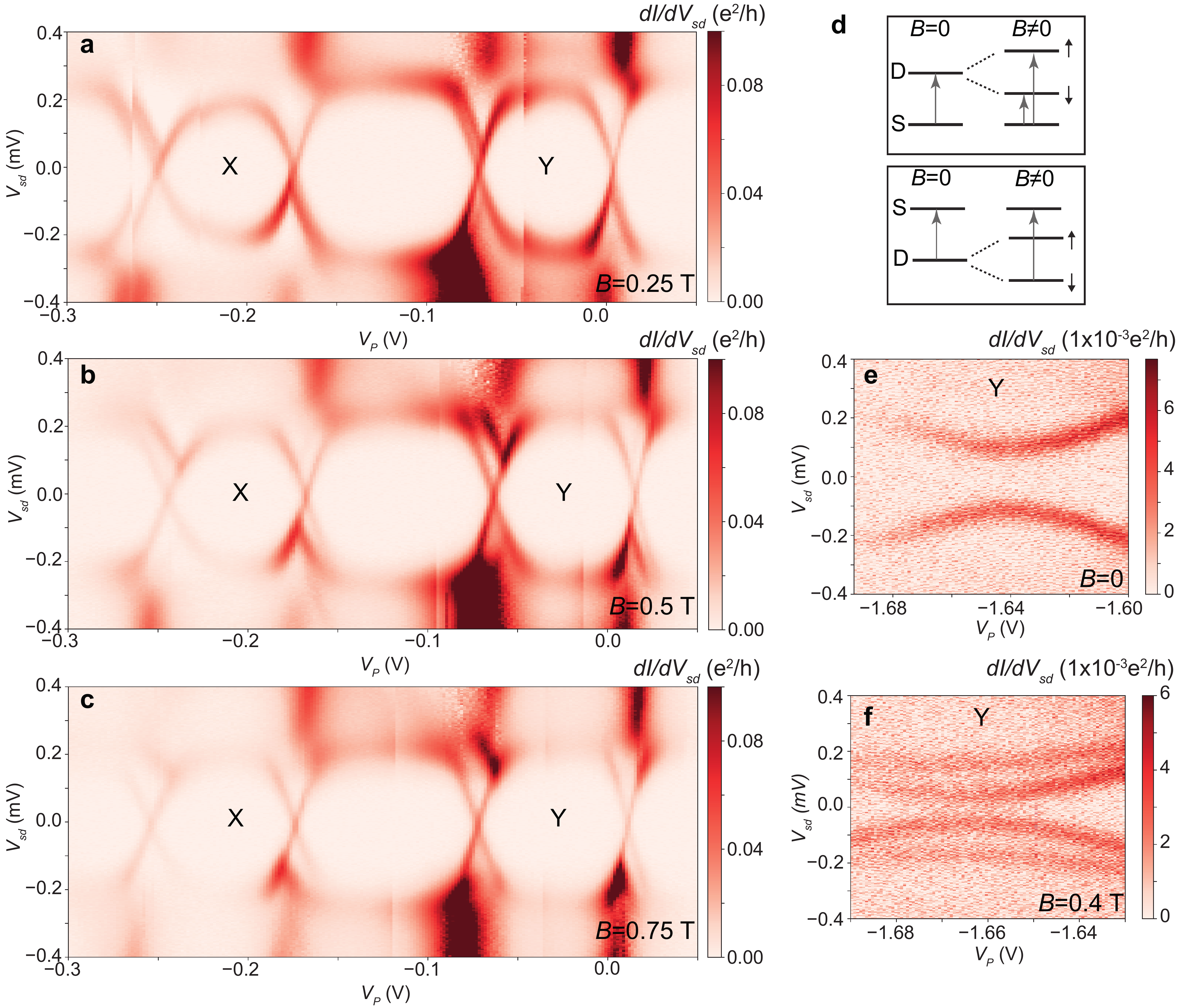}
\caption{\textbf{Ground-state corroboration through the magnetic evolution of YSR peaks at T=22 mK.} (a-c) Differential conductance colormap of Fig.~\ref{Fig1}d at increasing magnetic field. The scale is saturated to highlight the Zeeman splitting. (d) Sketch of the Zeeman splitting and allowed excitations of YSR states. (e,f) Zeeman splitting of the Y charge state when tuned into a singlet ground state.}
\label{Fig8}
\end{figure*}

We fitted $dI/dV_{sd}(V_{sd})$ curves to Gaussians
\begin{equation}
\frac{d I }{dV_{sd}}=A_+e^{-\frac{(V_{sd}-V_{sd+})^{2}}{2c_+^{2}}}+A_-e^{-\frac{(V_{sd}+V_{sd-})^{2}}{2c_-^{2}}}
\end{equation}
 where $A_+$ ($A_-$) represents the height of the positive (negative) bias peak, $V_{sd+}$ ($V_{sd-}$) represents the position of the positive (negative) bias peak, and $\approx 2.35c_+$ ($\approx 2.35c_-$) represents the width of the positive (negative) bias peak. The fit is good up to the quasiparticle continuum, where the peaks lose weight to it. The fits were done below a temperature at which it was not possible to distinguish the edge of the gap $\Delta$, or below a temperature at which the two YSR peaks merged into one (whichever was the lowest). This temperature limit varied within datasets. Across one dataset, we kept fixed the bias range in which the fit was performed.
 Figures \ref{Fig7}a-d show a typical example of the same pair of YSR peaks at different temperatures. The fit quality does not deteriorate with an increase in temperature. To verify the robustness of the parameters extracted from the fit, we compare in Figs.~\ref{Fig7}f,g the maximum of the peak and its position to the peak height and peak position extracted from the fit. Due to data noise (in Fig.~\ref{Fig7}d, fluctuations in conductance at the positive-bias YSR peak with respect to the Gaussian fit are $3 \times 10^{-3} e^2/h$), these two values are slightly different, but follow the same trend.

\subsection*{Gap hardness}

A soft superconducting gap is expected to provide additional relaxation channels which should result in extrinsic YSR peak broadening leading to a Lorentzian-shaped peak \cite{Martin2014Sep}. In our two devices, the gap is hard as evidenced by the subgap conductance suppression, while the experimentally extracted YSR peak lineshape is Gaussian. Fig.~\ref{Fig9} shows $dI/dV_{sd}(V_{sd})$ traces of the gap in linear and logarithmic scale measured in deep Coulomb blockade in the regime $\Gamma_N$, $\Gamma_S \ll U$, in which YSR and Kondo physics are suppressed. The FWHM of the positive-bias gap singularity is 50 $\mu$V for the etched device, and 40 $\mu$V for the shadowed one. These numbers are slightly smaller than the FWHM of the narrowest positive-bias YSR peak measured at e-h symmetry points in each device, which corresponds to 65 $\mu$V for the etched device, and 50 $\mu$V for the shadowed one.

\subsection*{Variation of the gap against temperature}

A decrease of the gap against temperature can produce motion of the position of the YSR peaks within the gap in the opposite direction as observed in the experiment \cite{Zitko2016May,Liu2019May}. In both the etched and shadowed devices we observe a decrease of the gap of no more than 5\%. In both cases, we determined this through the temperature dependence of the YSR peaks in the charge state of singlet ground state next to one of the examined doublets. In even charge states, the S-QD-N junction behaves effectively as a co-tunnelling junction without subgap YSR excitations, making this procedure feasible~\cite{Koerting2010Dec}. Figure~\ref{Fig10} shows that the $\Delta$ peaks remain constant in bias from 22 mK to 590 mK, despite losing weight. Note that the gap progressively fills with quasiparticle density of states as the temperature is increased.

\subsection*{Determination of the ground state by Zeeman-split spectroscopy}

As seen before in Refs.~\cite{lee2014spin,grove2017yu} and explained schematically in Fig.~\ref{Fig8}d, in a finite magnetic field the states of singlet ground state show two excitations corresponding to two spin-resolved excited doublets. This provides a way to distinguish them from states of doublet ground state, which show only one excitation. We verified the ground state of the charge states X and Y to which the datasets of Figs.~\ref{Fig1} to \ref{Fig5} correspond by observing the Zeeman splitting of the YSR peaks in an external magnetic field $B$. Figures \ref{Fig8}a-c show that the X and Y loops of doublet ground state expand with $B$ without any visible peak splitting. This growth is due to doublet splitting, as the energy of the spin-down doublet state is decreased. However, adjacent charge states of singlet ground state to the left and right of X and Y show peak splitting with $B$, with peaks splitting parallel-wise to their edges. We also corroborated the ground state of the Y charge state once it was tuned into a singlet ground state. Figures \ref{Fig8}e,f show splitting with magnetic field of the characteristic YSR peaks of singlet ground state~\cite{lee2014spin,grove2017yu}. Due to the lower critical field of the shadowed sample, this verification could not be performed for the datasets shown in Fig.~\ref{Fig6}.

\section*{Acknowledgements}

The authors are grateful to Dr.~Rok \v{Z}itko, Dr. Jens Paaske and Gorm Steffensen for fruitful discussions and to Mikelis Marnauza and Dags Olsteins for experimental assistance. The project received funding from the European Union’s Horizon 2020 research and innovation programme under the Marie Sklodowska-Curie grant agreement No.~832645. We acknowledge additional financial support from the Carlsberg Foundation, the Independent Research Fund Denmark, QuantERA ’SuperTop’ (NN 127900), Villum Foundation project No.~25310, the Danish National Research Foundation and the Sino-Danish Center for Education and Research. P.~K. acknowledges support from Microsoft and the ERC starting Grant No.~716655 under the Horizon 2020 program.

\section*{Author contributions statement}

J.C.E.S., K.G.R. and J.N. conceived the experiments, J.C.E.S., A.V. and V.S. conducted the experiments, T.K., P.K. and J.N. developed the nanowires. All authors reviewed the manuscript. 

\section*{Additional information}

The authors declare no competing interests. All data needed to evaluate the conclusions in the paper are present in the paper. Additional data related to this paper may be provided upon request on and after the publication date.

\end{document}